\documentclass[a4paper,twocolumn,11pt,accepted=2023-09-28]{quantumarticle}
\pdfoutput=1
\usepackage[numbers,sort&compress]{natbib}
\usepackage[utf8]{inputenc}
\usepackage[english]{babel}
\usepackage[T1]{fontenc}

\usepackage{tikz}
\usepackage{lipsum}

\usepackage{amsmath,amsthm,amsfonts,amssymb,mathrsfs,mathdots,graphicx,xcolor,times,xfrac,mathtools,enumerate,xr,subfigure,bbm,verbatim,comment,dsfont,soul,array,multirow}
\usepackage[T1]{fontenc}

\usepackage{color}
\usepackage[unicode=true,bookmarks=true,bookmarksnumbered=false,bookmarksopen=false,breaklinks=false,pdfborder={0 0 1}, backref=false,colorlinks=true]{hyperref}
\newcommand{\ket}[1]{\ensuremath{\left|\right.\!{#1}\!\left.\right\rangle}}
\newcommand{\bra}[1]{\ensuremath{\left\langle\right.\!{#1}\!\left.\right|}}

\newcommand{\subtiny}[3]{\ensuremath{_{\hspace{#1 pt}\protect\raisebox{#2 pt}{\tiny{$ #3$}}}}}
\newcommand{\suptiny}[3]{\ensuremath{^{\hspace{#1 pt}\protect\raisebox{#2 pt}{\tiny{$ #3$}}}}}

\DeclarePairedDelimiter{\norm}{\lVert}{\rVert}
\newcommand{\floors}[1]{ \left\lfloor  #1 \right\rfloor}

\begin{document}

\title{On the role of entanglement in qudit-based circuit compression}

\author{Xiaoqin Gao}
\affiliation{Department of Physics, University of Ottawa, Advanced Research Complex, 25 Templeton Street, K1N 6N5, Ottawa, ON, Canada}
\affiliation{Institute for Quantum Optics and Quantum Information -- IQOQI Vienna, Austrian Academy of Sciences, Boltzmanngasse~3, 1090 Vienna, Austria}
\orcid{0000-0001-5043-1122}
\author{Paul Appel}
\affiliation{Institute for Quantum Optics and Quantum Information -- IQOQI Vienna, Austrian Academy of Sciences, Boltzmanngasse~3, 1090 Vienna, Austria}
\orcid{0000-0001-8319-1321}
\author{Nicolai Friis}
\affiliation{Atominstitut, Technische Universit{\"a}t Wien, Stadionallee 2, 1020 Vienna, Austria}
\affiliation{Institute for Quantum Optics and Quantum Information -- IQOQI Vienna, Austrian Academy of Sciences, Boltzmanngasse~3, 1090 Vienna, Austria}
\orcid{0000-0003-1950-8640}
\author{Martin Ringbauer}
\affiliation{Universit{\"a}t Innsbruck, Institut f{\"u}r Experimentalphysik, Technikerstrasse 25, 6020 Innsbruck, Austria}
\orcid{0000-0001-5055-6240}
\author{Marcus Huber}
\affiliation{Atominstitut, Technische Universit{\"a}t Wien, Stadionallee 2, 1020 Vienna, Austria}
\affiliation{Institute for Quantum Optics and Quantum Information -- IQOQI Vienna, Austrian Academy of Sciences, Boltzmanngasse~3, 1090 Vienna, Austria}
\orcid{0000-0003-1985-4623}

\begin{abstract}
\noindent Gate-based universal quantum computation is formulated in terms of two types of operations: local single-qubit gates, which are typically easily implementable, and two-qubit entangling gates, whose faithful implementation remains one of the major experimental challenges since it requires controlled interactions between individual systems. To make the most of quantum hardware it is crucial to process information in the most efficient way. One promising avenue is to use higher-dimensional systems, qudits, as the fundamental units of quantum information, in order to replace a fraction of the qubit-entangling gates with qudit-local gates. Here, we show how the complexity of multi-qubit circuits can be lowered significantly by employing qudit encodings, which we quantify by considering exemplary circuits with exactly known (multi-qubit) gate complexity. We discuss general principles for circuit compression, derive upper and lower bounds on the achievable advantage, and highlight the key role played by entanglement and the available gate set. Explicit experimental schemes for photonic as well as for trapped-ion implementations are provided and demonstrate a significant expected gain in circuit performance for both platforms.
\end{abstract}

\maketitle

Quantum computation is a disruptive technology that has irrevocably changed the way that computation is envisioned. It holds the potential for addressing a wide range of computational challenges~\cite{Fedorov2022}, from factoring~\cite{shor1994algorithms} and database search~\cite{grover1996fast}, to applications in quantum machine learning~\cite{DunjkoBriegel2018,Cong2019}. Yet, current noisy intermediate-scale quantum (NISQ) devices~\cite{Preskill2018} are still far from addressing these applications at a practically relevant scale and early demonstrations of quantum advantages remain confined to algorithms that do not yet have clearly identifiable broad applications~\cite{arute2019quantumshort, zhu2021quantumshort}. The primary obstacle for today's quantum computers, which now feature 10s to 100s of qubits \cite{Bernien-Lukin2017, Zhang-Monroe2017, Otterbach-Rigetti2017, WangEtAl2018, FriisMartyEtal2018,  BradleyEtAl2019, PogorelovEtAl2021, MooneyWhiteHillHollenberg2021b} remains noise and decoherence, which limits the number of entangling operations and therefore the achievable circuit depth. Hence, for current as well as future quantum computers, efficiency at the circuit level will be key to getting the most out of these devices.\\[-1mm]

Fortunately, there is a lot of unused potential in current quantum devices, which tend to use only a small fraction of the available Hilbert space. Indeed, control over the inherently high-dimensional Hilbert space has been demonstrated in all major quantum technology platforms~\cite{Wang2018,Morvan2020,Ringbauer2021,Chi2022}, motivating the exploitation of a new paradigm of quantum computing based on $d$-dimensional \emph{qudits}, rather than qubits. Compared to their two-level counterpart, qudit architectures offer much richer coherence~\cite{Ringbauer2017} and entanglement structures~\cite{Kraft2018}, which can be exploited for efficient quantum information processing~\cite{lanyon2009simplifying,nikolaeva2021efficient,KiktenkoNikolaevaXuShlyapnikovFedorov2020,WangHuSandersKais2020} and improved quantum error correction~\cite{Watson2012,Campbell2014}. Since entangling operations tend to be the bottleneck in current quantum devices, the efficiency of a quantum computation, or the complexity of a quantum circuit, is traditionally measured by counting the number of entangling operations~\cite{haferkamp2022linear}. While this is an incomplete picture, it serves as a good hardware-agnostic approximation, because the optimal circuit for a given quantum operation is elusive and highly dependent on the available gate set.\\[-1mm]

Here, we investigate \emph{qudit circuit compression}, as a way to simplify a given qubit circuit by rephrasing it as a qudit circuit. We achieve this in two steps: First, the qubits are partitioned into groups of equal size such that the number of gates within the groups is maximized. Each group is then interpreted as a qudit, turning entangling gates into local gates and thus reducing the overall entangling gate count by a combinatorial factor for which we find lower and upper bounds using a graph-based approach. Second, by considering an extended gate set including not only qubit-entangling gates, but also genuine qudit-entangling gates, the number of entangling gates in the resulting qudit circuit can be further reduced, even saturating the combinatorial lower bound. To showcase these two effects, we study the compression of exemplary qubit circuits under different gate sets.
We illustrate this gate compression with experimental details for two contemporary quantum technologies using qudits: photonic qudits encoded in orbital angular momentum and trapped ions with multiple addressable levels, showing that, already today, qubit circuits can be more efficiently compiled on qudit architectures.
Aside from algorithmic improvements, reducing the number of quantum information carriers tends to make the system experimentally easier to control, leading to improved performance. We provide general design principles, as well as upper and lower bounds for the possible reduction in the number of gates, exemplified with explicit constructions for photonic systems and trapped ions.\\[-1mm]

\begin{figure*}[htb!]
    \centering
	\includegraphics[width=0.98\textwidth,trim={0cm 8cm 0cm 0cm},clip]{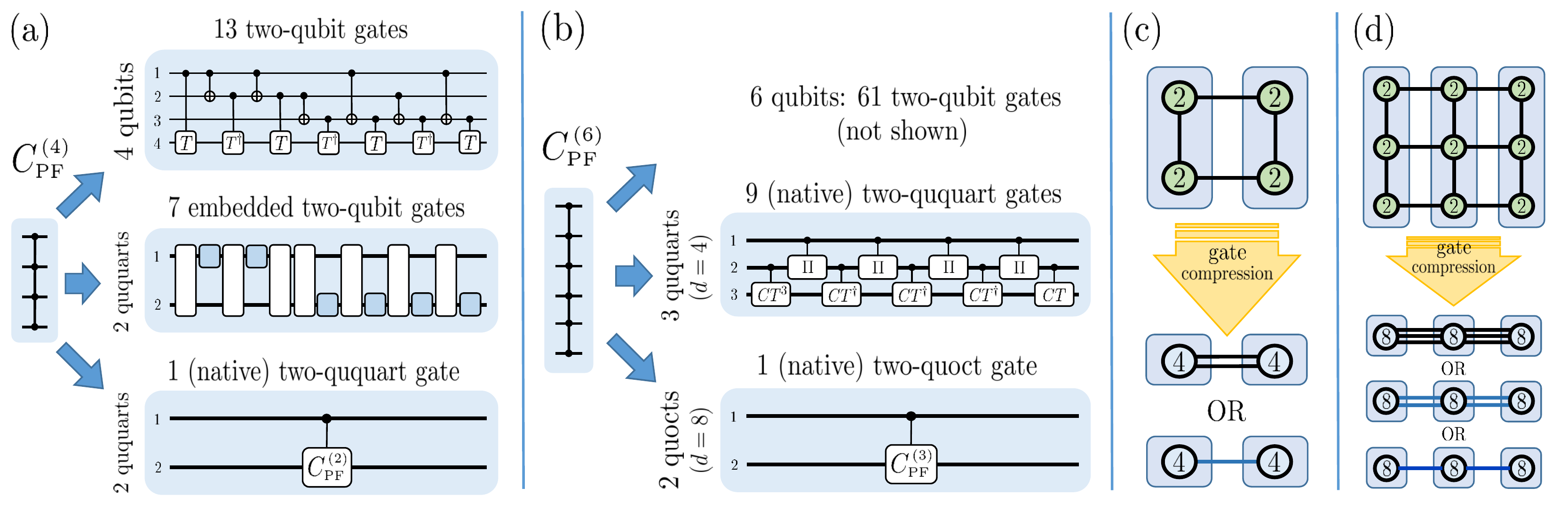}
    \caption{\textbf{Gate compression}.
    (a) \& (b)
    Compression of quantum circuits of $N$-qubit controlled phase-flip (CPF) gates $C\subtiny{0}{0}{\mathrm{PF}}\suptiny{1}{-1}{(N)}$. (a) The best known \cite{barenco1995elementary} decomposition of a $4$-qubit CPF gate $C\subtiny{0}{0}{\mathrm{PF}}\suptiny{1}{-1}{(4)}$ in terms of two-qubit gates requires $13$ entangling gates, i.e., $6$ CNOTs and $7$ controlled $T$ (or $T^{\dagger}$) gates (CT) with $T=\operatorname{diag}\{1,\exp(i\pi/4)\}$. By compressing the circuit to two qudits of dimension $4$, two quarts, only $7$ entangling gates are required in terms of embedded two-qubit gates, while $6$ previously non-local gates become local. However, the same operation $C\subtiny{0}{0}{\mathrm{PF}}\suptiny{1}{-1}{(4)}$ could also be realized by a single controlled-$C\subtiny{0}{0}{\mathrm{PF}}\suptiny{1}{-1}{(2)}=$CZ$=\operatorname{diag}\{1,1,1,-1\}$ gate on two ququarts, carried out if ququart $1$ is in the state $\ket{d-1}=\ket{3}$ corresponding to the two-qubit state $\ket{11}$.
    (b) Similarly, the most efficient decomposition of $C\subtiny{0}{0}{\mathrm{PF}}\suptiny{1}{-1}{(6)}$ for $6$ qubits requires $61$ two-qubit gates~\cite{barenco1995elementary}, whereas the compression to $3$ ququarts ($d=4$) requires (at most) $9$ (native) two-ququart gates, including adjoints and powers of controlled CT gates as well as controlled $\Pi$ gates, where $\Pi=\sum_{n=0}^{3}\ket{(n-1)_{\mathrm{mod}(3)}}\!\bra{n}$ is a permutation. Further compressing to two quocts ($d=8$), as little as $1$ controlled-$C\subtiny{0}{0}{\mathrm{PF}}\suptiny{1}{-1}{(3)}$ gate, conditioned on the quoct computational-basis state $\ket{d-1}=\ket{7}$ may be required. In both (a) and (b) the final decomposition of the $N$-qubit CPF gate into a single controlled-$C\subtiny{0}{0}{\mathrm{PF}}\suptiny{1}{-1}{(N/2)}$ gate requires a non-factorizable two-level entangling gate.
    (c) \& (d)
    Illustration of the effect of gate compression for the creation of a (quadratic) 2D cluster state consisting of four (c) and nine qubits (d), respectively. Cluster states are a particular type of graph state that can be schematically represented by vertices (here shown as circles carrying the local subsystem dimension $d$ as labels, 2 for qubits, 4 for ququarts, etc.) connected by lines representing entangling gates (controlled-$Z$ gates for qubits). By grouping (blue boxes) the qubits into pairs (c) or triples (d) the subsystem dimension becomes $d=4$ (ququart) and $d=8$ (quoct), respectively, but the number of non-local entangling gates can be reduced (c) from $4$ two-qubit gates to either $2$ embedded two-qubit gates or even $1$ single-qudit gate, and (d) from $12$ two-qubit gates to $6$ embedded two-qubit gates, $4$ two-qudit gates, or $2$ two-quoct gates, as is explained below.}
    \label{fig1:combinedfigure}
\vspace{-0.2cm}
\end{figure*}

\noindent\emph{Circuit compression.}\ 
Quantum circuits are built from a sequence of gate operations. At the lowest level of abstraction, circuit compression is hence concerned with compiling an $N$-qubit unitary to an $M$-qudit architecture ($M<N$), which we call \emph{gate compression}, see Fig.~\ref{fig1:combinedfigure}.
The task is to encode the qubit circuit into the qudit architecture in such a way that the maximal number of entangling gates in the qubit circuit manifests as local gates in the resulting qudit circuit. 
Importantly, the remaining entangling gates retain their qubit-entangling structure, in terms of maximally generated entanglement entropy, when embedded in the qudit Hilbert space. 
Consequently, this procedure always reduces the amount of entanglement needed (irrespective of the available gate set for qudits), by compressing non-local gates into local ones. Consider the example of a four-qubit circuit, where qubits 1 and 2 are encoded in one qudit, and qubits 3 and 4 in another. Now all non-local gates between the original qubits 1 and 2 as well as between qubits 3 and 4 are local in the respective qudits. The non-local gates between qubits that are now encoded in different qudits remain non-local and maintain their tensor-product structure, which is why we call them embedded qubit gates.

Importantly, while embedded qubit gates still create two-level entanglement (i.e., equivalent to a two-qubit gate in terms of entanglement entropy), they are not necessarily easily implementable in a qudit architecture. To see this, consider the example of a CNOT gate $U\subtiny{0}{0}{\mathrm{CNOT}}\suptiny{1}{0}{(c,t)}$ applied to qubits 2 (control) and 3 (target) in the above example of a four-qubit register, taking the form $\mathds{1}^{(1)} \otimes\, U\subtiny{0}{0}{\mathrm{CNOT}}\suptiny{1}{0}{(2,3)} \otimes \mathds{1}^{(4)}$. When qubits 1 and 2 are encoded in the first qudit, and 3 and 4 in the second, the resulting embedded version of the original gate would have to be a subspace-agnostic operation (i.e., applying the same operation on both subspaces pertaining to the encoded qubits 1 and 4). Already for a single-qudit example with the canonical encoding $\ket{0}=\ket{00},\ket{1}=\ket{01},\ket{2}=\ket{10},\ket{3}=\ket{11}$, we see that performing an operation of the form $\mathds{1} \otimes U$, which acts on the second encoded qubit only requires the application of the same $U$ to the subspaces $\{\ket{0},\ket{1}\}$ and $\{\ket{2},\ket{3}\}$ in order to realize the original tensor-product structure. While embedded qubit entangling gates still only create two-level entanglement, they are hence often not available natively. On the other hand, the qudit encoding enables new kinds of two-level entangling operations, which do not admit a tensor-product structure in the corresponding qubit circuit. Such gates provide new powerful tools for circuit compression, as discussed below, while also emphasizing the importance of the available gate set for qudit architectures. This again highlights the importance of considering the available gate set, rather than just qudit dimension, for efficient circuit compression.

While finding the ideal embedding of qubits into qudits is difficult, we can use weighted graphs to find a simplified representation of the circuit which in turn allows for powerful tools of graph theory to be employed. A weighted graph $G=(V,E)$ is a pair of a set of vertices $V=\{v_1,\dots,v_n\}$ and a set of edges $E=\{e_1,\dots,e_k\}$, where each edge $e_i= ((v_l,v_m),w_i)$ is a pair of vertices together with a weight $w_i$.
To encode the non-local properties of the quantum circuit into the graph, we associate each vertex with a qubit and draw an edge whenever a non-local gate is present between two qubits. The weight of each edge is determined by counting the number of non-local gates between the respective qubits. This graph representation simplifies the circuit by ignoring the gate order, but it works for both qubits and qudits. Using this graph representation, one can employ graph-partition algorithms~\cite{goldschmidt94, buluc_recent_2015} to divide the graph into subgraphs (partitions) in a way that maximizes the number of non-local gates within each partition. Specifically, we try to find a minimal $k$-cut, which involves identifying a set of edges $E'$ that, when removed, partitions the graph into $k$ parts while minimizing the sum of weights in $E'$. Note that this paradigm can cover a wide range of optimization goals, including unequal-size partitions or different cost functions. In particular, we highlight the importance of considering the structure of the original circuit, as well as the available forms of qudit entanglement in the target hardware to achieve optimal encoding of qubits into qudits.

\emph{Gate set optimization.}\ 
After we have reduced the \emph{width} of the circuit, we can further improve upon the circuit \emph{depth} by considering the extended gate set we can access for qudits. Recall that the embedded gates, resulting from circuit compression, still generate two-level entanglement and retain their original tensor-product structure. While such gates can be constructed from a universal set of qudit gates, there is much more untapped potential in the qudit Hilbert space, including gates that do not respect the tensor-product structure of the original qubit Hilbert space. By expanding the gate set, a significant reduction in circuit depth can thus be achieved, as illustrated in Fig.~\ref{fig1:combinedfigure}. However, the choice of gates for the expansion strongly depends on the chosen hardware platform.

\emph{Upper and lower bounds.}\ 
Using the above principles, we can derive upper and lower bounds on the \emph{compression ratio} $C_d$, which we define as the ratio of the number of entangling gates in the compressed circuit to the number of gates in the original circuit. To compute these bounds, we denote the weights of the original qubit graph by $\vec\omega$ and the weights of the qudit graph after compression by $\vec{\tilde{\omega}}$. The latter is obtained by dropping all components of the original vector that correspond to non-local gates between qubits that are encoded in the same qudit, and which thus become local, and then assigning all edges between qubits that are encoded in different qudits to the respective qudits in the new graph (and adding the corresponding weights). The upper (worst-case) bound is then obtained by comparing the sum of weights (i.e., the $1$-norm of $\vec\omega$) before and after compression, representing the scenario where no further gate optimization is possible. The lower bound, on the other hand, represents the scenario where gate optimization is able to fully exploit native qudit gates to realize the full qudit-entangling operation with a single gate operation. This bound is hence determined by the number of non-zero weights in $\vec{\tilde{\omega}}$. By normalizing these bounds using the number of gates in the initial circuit, $\norm{\vec{ w}}_1$, we can establish the compression-ratio bounds
\begin{equation}
    \frac{\norm{\vec{\tilde w}}_0}{\norm{\vec{w}}_1} \leq C_d \leq  \frac{\norm{\vec{\tilde w}}_1}{\norm{\vec{w}}_1} ,
\end{equation}
where, $\norm{\cdot}_0$ and $\norm{\cdot}_1$ refer to the $l_0$ (quasi-)norm and the $l_1$ norm, respectively. The span between the lower and upper bounds indicates potential improvements based on an appropriate gate set. In the following, we will provide examples saturating both the lower and the upper bound.

\emph{Controlled phase gates.}\ 
As a key example, we will now study $N$-qubit controlled phase-flip gates $C\subtiny{0}{0}{\mathrm{PF}}\suptiny{1}{-1}{(N)}$ (i.e., phase gates with a $\pi$ phase shift). These gates are central elements in quantum computing, for example, as a key part of Grover's search algorithm \cite{grover1996fast}. It can be written as follows,
\begin{equation}
\begin{aligned}
C\subtiny{0}{0}{\mathrm{PF}}\suptiny{1}{-1}{(N)} = \mathds{1}^{\otimes N}+\ket{1}\!\bra{1}^{\otimes (N-1)}\otimes(Z-\mathds{1}),
\label{eq:equation}
\end{aligned}
\end{equation}
where $\mathds{1}$ is the identity operator and $Z$ denotes the single-qubit Pauli-$z$ gate. Due to its central role, the decomposition of $C\subtiny{0}{0}{\mathrm{PF}}\suptiny{1}{-2}{(N)}$ is often used as a benchmark for comparing circuit decompositions and gate sets. Importantly, provably optimal decompositions into two-qubit gates and local operations are known for 3 to 8 qubits~\cite{barenco1995elementary}. As we noted above, the optimality of a quantum circuit depends on the metric used. However, even when the metric is fixed, finding an efficient way to realize the circuit is highly non-trivial.
For example, for a four-qubit controlled phase flip $C\subtiny{0}{0}{\mathrm{PF}}\suptiny{1}{-1}{(4)}$, the most efficient decomposition known, believed to be optimal, requires 13 two-qubit gates~\cite{barenco1995elementary}, as shown in Fig.~\ref{fig1:combinedfigure}~(a). The same circuit can be realized in a qudit system by embedding two qubits each into two ququarts. There are 3 ways to achieve such a partition, and the best choice turns 6 entangling gates into local gates, leaving 7 embedded two-qubit gates between the two ququarts. If we consider also ququart gates in our gate set, this can further reduce the non-local gate count to just a single gate between the two ququarts, which turn out to generate only two-level entanglement, as shown in Fig.~\ref{fig1:combinedfigure}~(a). In case $N>4$, we can cut $N$ qubits into $k$ equal parts by using a $2^{N/k}$-dimensional qudit in each part and achieve similar improvements. For the six-qubit controlled phase-flip gate ($C\subtiny{0}{0}{\mathrm{PF}}\suptiny{1}{-1}{(6)}$), for example, 61 two-qubit gates are needed in the best known (again, believed to be optimal) decomposition~\cite{barenco1995elementary}. Encoding the same circuit into 3 ququarts reduces the requirement to 9 two-ququart gates, including four two-ququart controlled-shift gates and five two-ququart controlled phase gates \cite{bullock2006efficient}, see Fig.~\ref{fig1:combinedfigure}~(b). Going further by encoding the gate into two qudits of dimension $d=8$ (quocts), the gate can again be realized with a single two-level entangling qudit gate.\\[-1mm]

\emph{Graph states and the saturation of graph partitioning bounds.--}
Graph states are another highly relevant example. For a set of edges $E=\{e\}$, they can always be created by applying controlled-$Z$ gates across all edges on computational-superposition product states, i.e., $|G\rangle:=\prod_{e\in E} CZ_e|+\rangle^{\otimes n}$ with $|+\rangle=\bigl(|0\rangle+|1\rangle\bigr)/\sqrt{2}$. As all $CZ_e$ commute, it directly follows that, with an unrestricted entangling power of the gate set, the lower bound $\norm{\vec{\tilde w}}_0$ can be saturated. If the gate set is restricted to two-level entangling gates, however, the fact that a high amount of entanglement may be generated if multiple edges cut across a bipartition directly implies that the upper bound needs to be observed with $\norm{\vec{\tilde w}}_1 $ embedded qubit gates.

Consider, for example, a 4-qubit quadratic 2D cluster state as illustrated in Fig.~\ref{fig1:combinedfigure}~(c). Due to the symmetry of the state the partition is irrelevant and we can thus combine the first and second qubit as well as the third and fourth qubit to a qudit each without loss of generality. This reduces the number of non-local gates from $4$ to $2$, which saturates the upper bound $\norm{\vec{\tilde w}}_1$. Now, we can use gate optimization to reduce the number to $1$ by combining the two non-local gates between the qudits to a single genuine qudit gate, saturating the lower bound $\norm{\vec{\tilde w}}_0$. This example demonstrates that the commutativity of the entangling gates across compressed partitions as well as the maximum amount of entanglement generated across these partitions are crucial parameters that determine how well a qubit circuit can be compressed when compiled on a qudit architecture and which gate set would be required for the qudits.

\emph{Photonic implementation.--}
Photonic systems are excellent candidates for gate compression. Current developments enable increased control over higher-dimensional degrees of freedom by manipulating a photon's polarization, spatial profile, temporal profile, or frequency, either separately or simultaneously. This enables the encoding of multiple bits into a single photon, as is routinely done in entanglement-based quantum communication \cite{Sit:17,AchatzEtAl2022,Herrera-ValenciaSrivastavPivoluskaHuberFriisMcCutcheonMalik2020}. Local unitary operations are easily done within a certain degree of freedom, such as spatial manipulation through multi-plane light conversion \cite{Brandt:20}, frequency manipulation \cite{kues2017chip,CabrejoPonceMarquesMunizHuberSteinlechner2022}, or between different degrees of freedom, for example, using polarizing beam splitters to couple path and polarization. For instance, high-dimensional Pauli $X$- and $Z$-gates, which are parts of higher-dimensional universal gate sets, have recently been implemented in a number of ways \cite{asadian2016heisenberg, babazadeh2017high, wang2015quantum, gao2019arbitrary, Brandt:20}. While local gates can often be performed with near-unit efficiency and fidelity, entangling gates remain the Achilles heel of photonic information processing, as entangling two photons can be achieved probabilistically at best, leading to an exponential decrease in success probability with the number of entangling gates. 

The latter aspect is where high-dimensional encodings can become particularly useful, as more information can be processed at higher fidelity and with potentially much higher success probability. Recent developments already show promising results~\cite{muthukrishnan2000multivalued, erhard2018twisted}, including the implementations of high-dimensional multi-partite quantum gates~\cite{gao2019computer} and of the SUM gate (a high-dimensional controlled-$X$ gate) in the time and frequency degrees of freedom of photons~\cite{imany2019high}. Here, we propose a scheme for a photonic two-qudit entangling gate that grows only logarithmically in complexity with dimension and achieves a constant success probability of $1/4$ that is independent of the qudit dimension, see Appendix~\ref{sec:appendix A} for details.

\emph{Trapped-ion implementation.--} Trapped ions are among the leading platforms for quantum information processing~\cite{Bermudez2017}, where the electronic energy levels of each ion naturally provide a high-dimensional Hilbert space. Recently it was shown that such a system can be operated as a universal qudit quantum processor up to dimension 7~\cite{Ringbauer2021}. The qudit-gate set used in this demonstrations consisted of arbitrary local gates and two-qubit CNOT gates embedded in a qudit Hilbert space. Compared to a standard qubit CNOT, the embedded version exhibits error rates larger by roughly a factor of 2, independent of the Hilbert-space dimension. Beyond this basic gate set, it has been shown that both dominant gate mechanisms in this platform, the M{\o}lmer-S{\o}rensen gates~\cite{Low2020} and light-shift gates~\cite{Hrmo2022}, can be generalized to achieve genuine qudit entanglement. A first experimental realization of the latter demonstrated the generation of genuine qudit entanglement in a scalable fashion and with highly competitive error rates~\cite{Hrmo2022}.

Compiling the example of Fig.~\ref{fig1:combinedfigure}~(a) and using state-of-the-art error rates for trapped-ion quantum processors of about 0.01 per qubit CNOT gate~\cite{Bermudez2017}, a rough estimate suggests that the implementation of the four-qubit $C\subtiny{0}{0}{\mathrm{PF}}\suptiny{1}{-1}{(4)}$ gate could achieve an error rate on the order of 0.12 with the standard two-level decomposition using 13 two-qubit entangling gates. Curiously, while it is known on the one hand that enlarging the Hilbert space locally (i.e., encoding 4 qubits into 4 qudits) can reduce the required number of gates quadratically~\cite{lanyon2009simplifying}, this gain is offset almost exactly in the 4-qubit case by the factor of 2 increased error rates incurred in the experimental implementation~\cite{Ringbauer2021}. On the other hand, when two qubits each are encoded in a qudit of dimension 5 (4 computational levels and 1 auxiliary level), the required number of two-level entangling gates drops to 1 (albeit with a rotation angle equivalent to 4 qubit gates), achieving an estimated error rate of 0.04, see Appendix~\ref{sec:appendix B} for the circuit. Curiously, this is an example where it is optimal to use a two-level entangling gate in the qudit circuit, which is not an embedded qubit gate.

\emph{Conclusion.--}
We have explored two ways in which higher-dimensional architectures are universally beneficial to quantum computing. First, using gate compression, we can cut down on the amount of entanglement needed for a specific $N$-qubit circuit, and second, by exploiting the added capabilities of qudit systems in the form of richer gate sets, we can further reduce the number of non-local gates required to generate that entanglement. This highlights the key role played by entanglement and the available gate set in efficient qudit QIP. As every higher-dimensional gate can be achieved by any universal gate set applied a number of times, the advantage from larger gate sets is constant in the number of qubits. Similarly, even the most efficient partitioning of the qubit circuit leads to only a constant advantage. Such constant improvements, however, can make the difference between feasibility and failure.

Another important aspect is the potential breakdown of conventional wisdom regarding easily implementable local gates versus hard non-local gates. Once qudit dimensions get sufficiently large, the performance gap between local and non-local gates might change. Moreover, gate performance typically degrades somewhat with system size, providing another motivation for reducing the number of quantum-information carriers and making more efficient use of available resources. Finally, we emphasize that a case-by-case evaluation of the actual trade-offs is critical to finding the optimal dimensionality for a given problem and hardware platform.

Our examples provide a promising first step, showing that qudit encodings can lead to a significant reduction in gate count. Hence, this approach can greatly increase the utility of current and future quantum hardware, using only degrees of freedom that are already present in today's quantum technology. Indeed, various quantum-computing platforms have demonstrated qudit control with ever-increasing performance. Both gate compression and gate-set optimization will be central tools for making the most of the next generation of high-dimensional quantum processors, harnessing the full potential of physical quantum information carriers.

\emph{Acknowledgments.--}	
We are indebted to Mateus Ara\'ujo for discussion and input during early stages of this work, and we thank him for his Krapfen.
We acknowledge financial support from the Austrian Science Fund (FWF) through the START project Y879-N27, the stand-alone project P~31339-N27, and the stand-alone project P~36478-N funded by the European Union – NextGenerationEU, as well as by the Austrian Federal Ministry of Education, Science and Research via the Austrian Research Promotion Agency (FFG) through the flagship project FO999897481 funded by the European Union – NextGenerationEU. This project has received funding from the European Union's Horizon 2020 research and innovation programme under the Marie Sk{\l}odowska-Curie grant agreement No 840450. X.G. acknowledges the support of Joint Centre for Extreme Photonics (JCEP). M.H. acknowledges funding from the European Research Council (Consolidator grant ’Cocoquest’ 101043705). M.R. acknowledges funding from the European Research Council (Starting Grant 'QUDITS' 101039522) and the European Union under the Horizon Europe Programme – Grant Agreement 101080086 – NeQST. Views and opinions expressed are however those of the author(s) only and do not necessarily reflect those of the European Union or the European Research Council. Neither the European Union nor the granting authority can be held responsible for them.


\bibliographystyle{apsrev4-1fixed_with_article_titles_full_names_new}
\bibliography{refs}


\newpage
\clearpage
\onecolumngrid

\hypertarget{sec:appendix}
\appendix

\section{Photonic Implementation with the orbital angular momentum}\label{sec:appendix A}

To implement quantum circuits with high-dimensional quantum gates, we propose a general experimental scheme for two-qudit CPF gates in the orbital angular momentum (OAM) of two photons. The scheme succeeds with a probability of $\tfrac{1}{4}$, irrespective of the encoding dimension, but requires an auxiliary two-qubit Bell state. Hence, this scheme does not overcome some of the fundamental limitations that all photonic computing suffers from but carries the potential to increase the number of gates that can be realized with a given set of resources, as well as exploring gates in one of the many potential degrees of freedom that can be harnessed in single-photons.

In order to implement an efficient photonic circuit for $C\subtiny{0}{0}{\mathrm{PF}}\suptiny{1}{-1}{(N)}$ in the OAM degree-of-freedom via a qudit encoding, high-dimensional quantum gates with high success probability are required. These include single-qudit gates, which have been well developed for the OAM of a single photon, i.e., Pauli $X$- \cite{babazadeh2017high, gao2019arbitrary, Brandt:20} and $Z$-gates \cite{wang2015quantum},
and non-local two-qudit quantum gates, which include two-qudit quantum controlled-phase gates or two-qudit quantum controlled-cyclic (permutation) gates. All of these non-local quantum gates are conditioned on the highest state $\ket{d-1}$ of the control qudit with computational basis $\{\ket{m}\}_{m=0,1,\ldots,d-1}$.

\begin{figure*}[b!]
    \centering
    \includegraphics[width=0.92\textwidth,trim={0cm 2.8cm 0cm 3,2cm},clip]{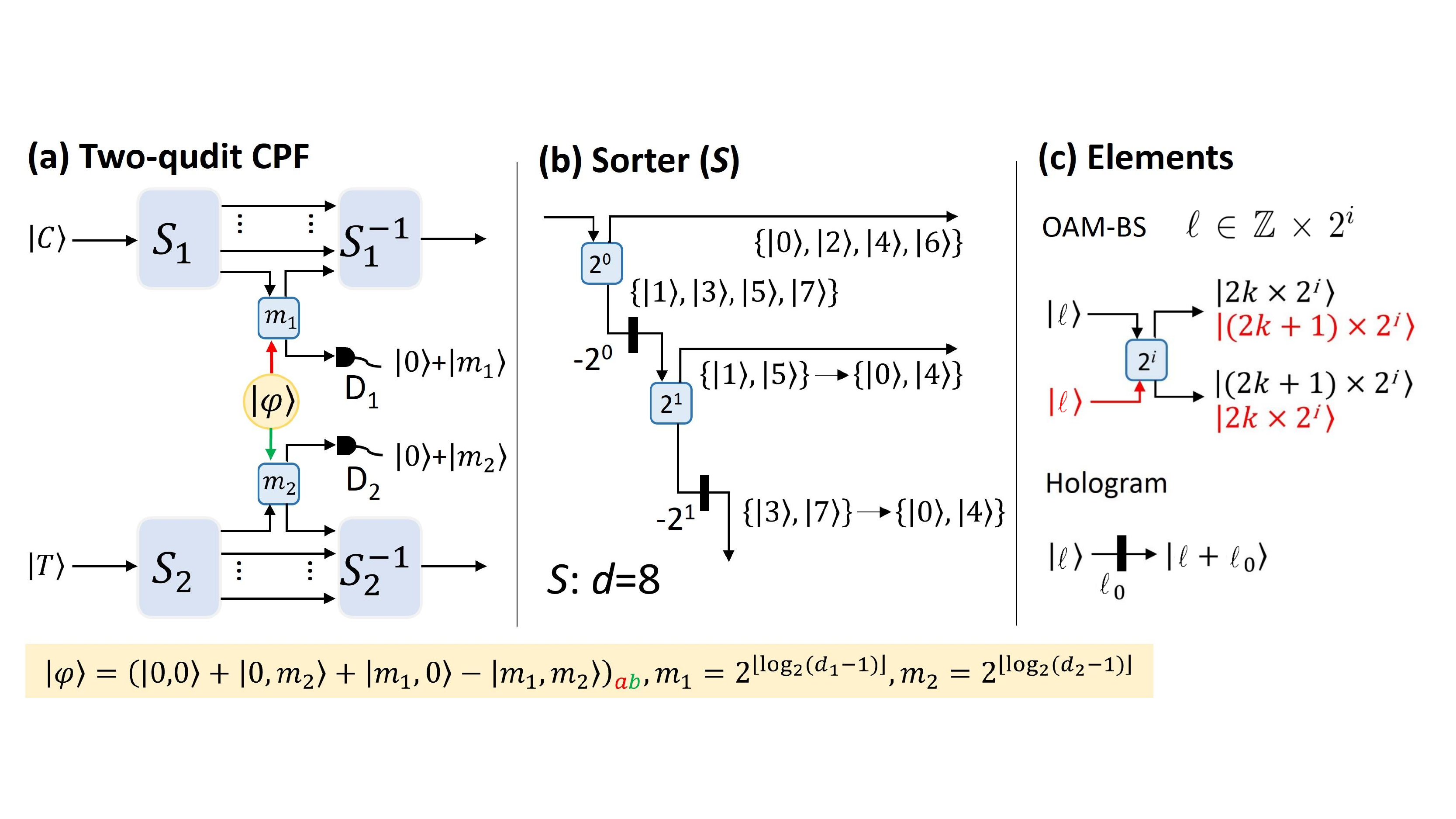}
    \caption{\textbf{Experimental scheme for realizing a two-qudit CPF gate in the OAM degree-of-freedom of two photons}.
    \textbf{(a)} The setup has three main parts representing the two subsystems in question, control (upper part) and target (lower part), as well as an auxiliary state $\ket{\varphi}$ in a two-qubit subspace spanned by two OAM modes ($\ket{0}$ and $\ket{m_{1}}$, as well as $\ket{0}$ and $\ket{m_{2}}$, respectively) each of two additional photons, here shown for arbitrary local dimension $d$.
    The control and target parts each consist of single-photon input states $\ket{C}$ and $\ket{T}$ that are fed into OAM sorters $S_{1}$ and $S_{2}$, respectively, which split pairs of the different OAM modes into orthogonal spatial modes. There are $\floors { \log_2 {(d_{1/2}-1)} }+1$  outputs from the sorter $S_{1/2}$. For each of these parts, the paths corresponding to the pair of OAM modes $\ket{d_{1/2}-1-2^{\floors { \log_2 {(d_{1/2}-1)} }}}$ and $\ket{d_{1/2}-1}$, respectively, are each combined with one of the photons of $\ket{\varphi}$ (normalization not shown) on additional OAM beam splitters (OAM-BS) $m_{1}=2^{\floors { \log_2 {(d_{1}-1)} }}$ and $m_{2}=2^{\floors { \log_2 {(d_{2}-1)} }}$ respectively. One of the output paths of each OAM-BS is then projected onto an equally weighted superposition of $\ket{0}$ and $\ket{m_{1/2}}$, while the remaining output paths of the OAM-BSs are fed back into a final inverse OAM-sorter operation $S^{-1}_{1/2}$ that recombines the different paths into a single path.
    \textbf{(b)} Example for OAM sorter $S$ in dimension eight ($d=8$) realized by a combination of two OAM-BSs [labeled by $2^i=1$ and $2^i=2$ with $i=0$ and $i=1$, respectively, as shown in \textbf{(c)}] and two holograms [for $\ell_{0}=-1$ and $\ell_{0}=-2$ as shown in \textbf{(c)}]. The first OAM-BS ($2^i=1$, $i\in \mathbb{Z}^{0+}$) separates the modes $\ket{0}$, $\ket{2}$, $\ket{4}$, and $\ket{6}$ from $\ket{1}$, $\ket{3}$, $\ket{5}$, and $\ket{7}$. The first hologram ($\ell_{0}=-1$) shifts the latter modes to $\ket{0}$, $\ket{2}$, $\ket{4}$, and $\ket{6}$, before the second OAM-BS ($2^i=2$) separates $\ket{0}$ and $\ket{4}$ (formerly $\ket{1}$ and $\ket{5}$) from $\ket{2}$ and $\ket{6}$ (formerly $\ket{3}$ and $\ket{7}$). The second hologram ($\ell_{0}=-2$) then maps  $\ket{2}$ and $\ket{6}$ to  $\ket{0}$ and $\ket{4}$, originally corresponding to $\ket{d-1-2^{\floors { \log_2 {(d-1)} }}}=\ket{3}$ and $\ket{d-1}=\ket{7}$.
    \textbf{(c)} Two basic elements: OAM beam splitters (OAM-BSs) and holograms. The label $i$ determines the sorting property of the OAM-BS. Holograms shift the azimuthal quantum number of the OAM modes fixed amount. For details about the structure of OAM-BS see Fig.~\ref{fig3:figure3}~(b).
    }
    \label{fig2:figure2}
\vspace{-0.2cm}%
\end{figure*}

\begin{figure}[h]
    \centering
    \includegraphics[scale=.6]{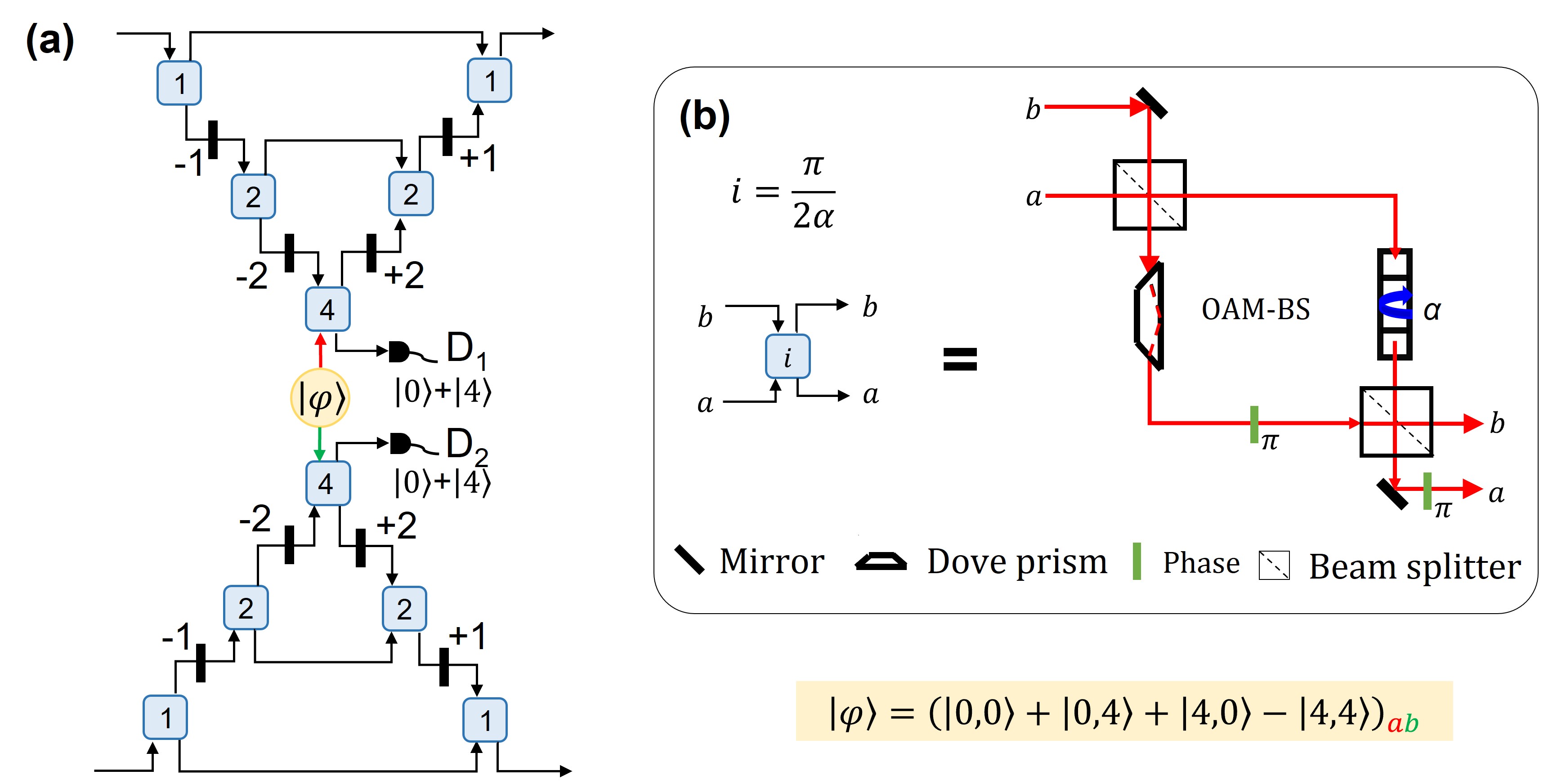}
    \caption{\textbf{Experimental scheme for two-qudit controlled phase flip in the OAM of two photons with eight dimensions}.
    (a) Experimental setup. 10 OAM-BSs (blue boxes) together with holograms (thick, black, vertical lines) are used in the setup.
    A two-qubit Bell state is necessary for facilitating the interaction between the two photons.
    (b) Structure of the OAM-BS.
    The OAM-BS works for sorting OAM modes, the sorting property is determined by $i$, which is related to the rotation angle $\alpha$ of the Dove prism.
     }
    \label{fig3:figure3}
\vspace{-0.2cm}%
\end{figure}

Here we are interested in performing two-qudit gates in the subspace spanned by the OAM modes $\{\ket{m}\}_{m=0,1,\ldots,d_{1}-1}$ and $\{\ket{m}\}_{m=0,1,\ldots,d_{2}-1}$ of two photons, where $d_{1}$ and $d_{2}$ are any (integer) dimensions. The main idea of our proposal for a potential experimental implementation is shown in Fig.~\ref{fig2:figure2}~(a).

The operation $S$ splits modes from one path into multiple paths, while $S^{-1}$ does the opposite, regrouping all modes into a single path. First, the input photon goes through $S$, which sends the highest mode $\ket{d-1}$ to its own path together with the mode $\ket{d-1-2^{\floors { \log_2 {(d-1)} }}}$ by using $\floors { \log_2 {(d-1)} }$ OAM-BSs and $\floors { \log_2 {(d-1)} } $ holograms. Subsequently, $\ket{d-1}$ and $\ket{d-1-2^{\floors { \log_2 {(d-1)} }}}$ are combined with a photon from the Bell state $\ket{\varphi}$ in an OAM-BS. Finally, the operation $S^{-1}$ recombines all modes into one single path. The gate works if two single-photon detectors click at the same time with probability of 1/4 (due to the auxiliary Bell state, normalization not shown). As an example, we show the operation $S$ for dimensions $d=8$ in Fig.~\ref{fig2:figure2}~(b). The highest mode $\ket{7}$ and mode $\ket{3}$ are split into their own paths and enter into an OAM-BS together with another photon (red) from the auxiliary state. Afterward, all modes are routed into one single path via $S^{-1}$. The operation $S$ consists of multiple OAM-BSs and holograms, as shown in Fig.~\ref{fig2:figure2}~(c). An OAM-BS is an interferometer containing two Dove prisms \cite{gonzalez2006dove}, see Fig.~\ref{fig3:figure3}. The operations $S$ and $S^{-1}$ features a high symmetry: $S^{-1}$ is a mirror reflection of $S$, with an inversion of the hologram values.

There are two basic elements in the setup: holograms and OAM beam splitters, each of which has a finite (below 1) fidelity, meaning an advantage can only materialise in high dimensions if one needs at most a logarithmic number of these elements for a specific gate. And indeed, the number $N(d_{1/2})$ of OAM-BSs scales logarithmically with $d_1$ and $d_2$:
\begin{align}
N(d_{1/2}) = 2 \times (\floors { \log_2 {(d_1-1)} }+\floors { \log_2 {(d_2-1)} }) +2,
\label{eq:equation2}
\end{align}
where $N(d_{1/2})$ is an upper bound obtained by studying a simple setup.
For certain circuits, we know the actual number can still be lower. In the case of two eight-dimensional qudits, the complete $C\subtiny{0}{0}{\mathrm{PF}}\suptiny{1}{-1}{(6)}$ can be achieved with 10 OAM-BSs, as shown in Fig.~\ref{fig3:figure3}.

Remarkably, the probability of success is 1/4, irrespective of the dimension. Therefore, the efficiency of the circuit increases significantly when using higher-dimensional gates. It is also interesting to implement such gates in other degrees of freedom of photons, such as path.

\section{Trapped ion implementation}\label{sec:appendix B}

\begin{figure*}[h]
    \centering
    \begin{minipage}{0.03\textwidth}
    \ \\
    \ \\
    \ \\
    (a)\\
    \ \\
    \ \\
    \ \\
    \ \\
    (b)
    \end{minipage}
    \begin{minipage}{0.95\textwidth}
    \includegraphics[width=\textwidth]{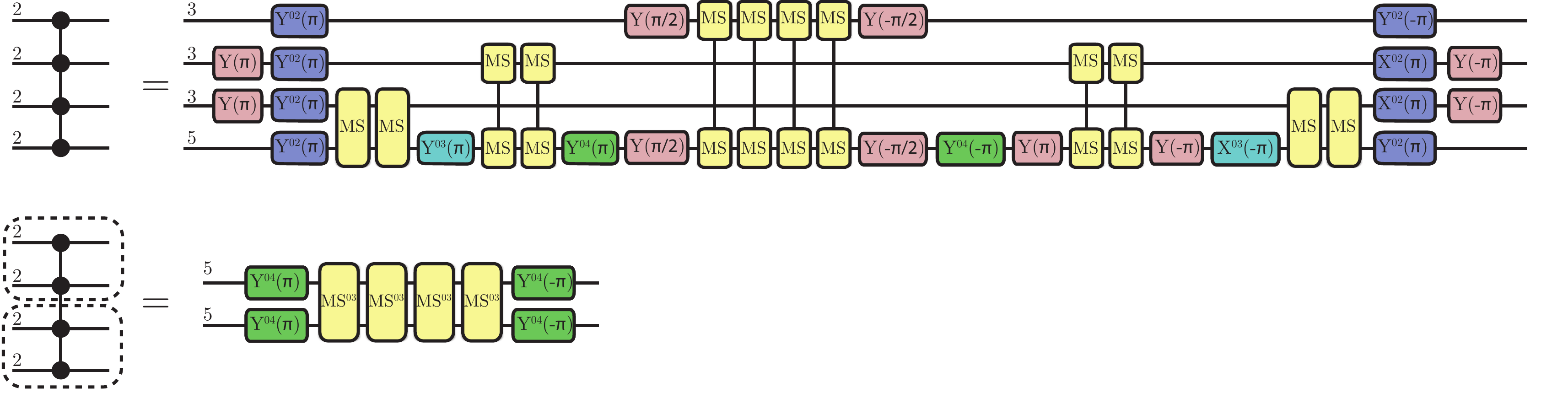}
    \end{minipage}
    \caption{\textbf{Trapped-ion circuits for implementing a four-qubit CPF gate $C\subtiny{0}{0}{\mathrm{PF}}\suptiny{1}{-1}{(4)}$}. Implementations are shown for (a) 4 qubits with auxiliary level, or (b) two qudits that encode 2 qubits each. In both cases only two-level entangling operations are used, where each yellow gate corresponds to a M{\o}lmer-S{\o}rensen (MS) gate with rotation angle $\pi/2$, which is maximally entangling for qubits. The local rotations are X or Y rotations with acting on the subspace indicated in the superscript (and color coding), where operations without superscript act on the 01-subspace of the original qubits. Note that, while using auxiliary levels enables a reduction to only 5 non-local gates, those gates require larger rotation angles and thus come with an increased experimental cost per gate. Even taking this into account the qudit-assisted circuit is expected to perform slightly better than a pure qubit circuit. In case (b) each ion occupies a 5-dimensional Hilbert space and encodes two qubits. Exploiting again the auxiliary level, only a single non-local gate is required.}
    \label{fig4:TIcircuits}
\end{figure*}

In Fig.~\ref{fig4:TIcircuits}, we provide a circuit for the implementation of a four-qubit CPF gate $C\subtiny{0}{0}{\mathrm{PF}}\suptiny{1}{-1}{(4)}$ using either four qubits, or 2 qudits.

\end{document}